\documentclass[11pt]{article}
\usepackage{a4}
\usepackage{natbib}

\usepackage[english]{babel}
\usepackage{algorithm}
\usepackage{algorithmic}
\usepackage{amsmath}
\usepackage{graphicx}
\usepackage{geometry}
\usepackage{times}
\usepackage{url}
\usepackage{multirow}
\begin{document}

\title{MissForest - nonparametric missing value imputation for
  mixed-type data}

\author{Daniel J. Stekhoven\,$^{\textsf{1,2,3,}}$\footnote{to whom
    correspondence should be addressed}~~and Peter
  B\"uhlmann\,$^{\textsf{1,3}}$\\
$^{\textsf{1}}$Seminar for Statistics, ETH Zurich, Zurich,
  Switzerland,\\$^{\textsf{2}}$Life Science Zurich PhD Program on Systems
  Biology of Complex Diseases,\\$^{\textsf{3}}$Competence Center for Systems
  Physiology and Metabolic Diseases, Zurich, Switzerland}

\date{}

\maketitle

\begin{abstract}
  {\bf Motivation:} Modern data acquisition based on high-throughput
  technology is often facing the problem of missing data. Algorithms
  commonly used in the analysis of such large-scale data often depend on a
  complete set. Missing value imputation offers a solution to this
  problem. However, the majority of available imputation methods are
  restricted to one type of variable only: continuous or categorical. For
  mixed-type data the different types are usually handled
  separately. Therefore, these methods ignore possible relations between
  variable types. We propose a nonparametric method which can cope with
  different types of variables simultaneously.

  {\bf Results:} We compare several state of the art methods for the
  imputation of missing values. We propose and evaluate an iterative
  imputation method (missForest) based on a random forest. By averaging
  over many unpruned classification or regression trees random forest
  intrinsically constitutes a multiple imputation scheme. Using the
  built-in out-of-bag error estimates of random forest we are able to
  estimate the imputation error without the need of a test set. Evaluation
  is performed on multiple data sets coming from a diverse selection of
  biological fields with artificially introduced missing values ranging
  from 10\% to 30\%. We show that missForest can successfully handle
  missing values, particularly in data sets including different types of
  variables. In our comparative study missForest outperforms other methods
  of imputation especially in data settings where complex interactions and
  nonlinear relations are suspected. The out-of-bag imputation error
  estimates of missForest prove to be adequate in all
  settings. Additionally, missForest exhibits attractive computational
  efficiency and can cope with high-dimensional data.

  {\bf Availability:} The \textsf{R} package \texttt{missForest} is freely
  available from \url{http://stat.ethz.ch/CRAN/}.
\end{abstract}

\begin{center}
\parbox{0.6\textwidth}{{\bf Pre-print version:} This article has been
  submitted to Oxford Journal's Bioinformatics$^\copyright$ on 3rd of May
  2011. Version 2 has been resubmitted on 27th of September 2011.}
\end{center}

\section{Introduction}\label{intro.sec}
Imputation of missing values is often a crucial step in data analysis. Many
established methods of analysis require fully observed data sets without
any missing values. However, this is seldom the case in medical and
biological research today.  The ongoing development of new and enhanced
measurement techniques in these fields provides data analysts with
challenges prompted not only by high-dimensional multivariate data where
the number of variables may greatly exceed the number of observations but also
by mixed data types where continuous and categorical variables are
present. In our context categorical variables can arise as any kind ranging
from technical settings in a mass spectrometer to a diagnostic expert
opinion on a disease state. Additionally, such data sets often contain
complex interactions and nonlinear relation structures which are
notoriously hard to capture with parametric procedures.

Most prevalent imputation methods, like $k$ nearest neighbours
(KNNimpute, \cite{troyanskaya01}) for continuous data, saturated multinomial
model (\cite{schafer97}) for categorical data and multivariate imputation
by chained equations (MICE, \cite{vanbuuren99}) for mixed data types
depend on tuning parameters or specification of a parametric model. The
choice of such tuning parameters or models without prior
knowledge is difficult and might have a dramatic effect on a method's
performance. Excluding MICE the above methods and the majority of other
imputation methods are restricted to one type of variable. Furthermore,
all these methods make assumptions about the distribution of the data or
subsets of the variables, leading to questionable situations,
e.g. assuming normal distributions.

The literature on mixed-type data imputation is rather scarce. Its first
appearance was in the developing field of multiple imputation brought up by
\cite{rubin78}. \cite{little85} presented an approach based on maximum
likelihood estimation combining the multivariate normal model for
continuous and the Poisson/multinomial model for categorical data. This
idea was later on extended in the book of \cite{little87}. See also
\cite{li88}, \cite{rubin90} and \cite{schafer97}. A more refined method to
combine different regression models for mixed-type data was proposed by
\cite{vanbuuren99} using chained equations. The conditional model in MICE
can be specified for the missing data in each incomplete
variable. Therefore no multivariate model covering the entire data set has
to be specified. However, it is assumed that such a full multivariate
distribution exists and missing values are sampled from conditional
distributions based on this full distribution (for more details see Section
\ref{methods.sec}).  Another similar method using variable-wise conditional
distributions was proposed by \cite{raghunathan01} called sequential
regression multivariate imputation. Unlike in MICE the predictors must not
be incomplete. The method is focussed on survey data and therefore includes
strategies to incorporate restrictions on subsamples of individuals and
logical bounds based on domain knowledge about the variables, e.g., only
women can have a number of pregnancies recorded.

Our motivation is to introduce a method of imputation which can handle any
type of input data and makes as few as possible assumptions about
structural aspects of the data. Random forest (RF, \cite{breiman01}) is
able to deal with mixed-type data and as a nonparametric method it allows
for interactive and nonlinear (regression) effects. We address the missing
data problem using an iterative imputation scheme by training a RF on
observed values in a first step, followed by predicting the missing values
and then proceeding iteratively. \cite{mazumder10} use a similar approach
for the matrix completion problem using a soft-thresholded SVD iteratively
replacing the missing values. We choose RF because it can handle mixed-type
data and is known to perform very well under barren conditions like high
dimensions, complex interactions and nonlinear data structures. Due to its
accuracy and robustness RF is well suited for the use in applied research
often harbouring such conditions. Furthermore, the RF algorithm allows for
estimating out-of-bag (OOB) error rates without the need for a test
set. For further details see \cite{breiman01}.

Here we compare our method with $k$-nearest neighbour imputation
(KNNimpute, \cite{troyanskaya01}) and the Missingness Pattern Alternating
Lasso (MissPALasso) algorithm by \cite{staedler10b} on data sets having
continuous variables only. For the cases of categorical and mixed type of
variables we compare our method with the MICE algorithm by
\cite{vanbuuren99} and a dummy variable encoded KNNimpute. Comparisons are
performed on several data sets coming from different fields of life
sciences and using different proportions of missing values.

We show that our approach is competitive to or outperforms the compared
methods on the used data sets irrespectively of the variable type
composition, the data dimensionality, the source of the data or the amount
of missing values. In some cases the decrease of imputation error is up to
50\%. This performance is typically reached within only a few iterations
which makes our method also computationally attractive. The OOB imputation
error estimates give a very good approximation of the true imputation error
having on average a proportional deviation of no more than
10~-~15\%. Furthermore, our approach needs no tuning parameter, hence, is
easy to use and needs no prior knowledge about the data.

\section{Approach}\label{approach.sec}
We assume ${\bf X} = ( {\bf X}_1, {\bf X}_2, \dots, {\bf X}_p)$ to be a
$n\times p$-dimensional data matrix. We propose using a random forest to
impute the missing values due to its earlier mentioned advantages as a
regression method. The random forest algorithm has a built-in routine to
handle missing values by weighting the frequency of the observed values in
a variable with the random forest proximities after being trained on the
initially mean imputed data set (\cite{breiman01}). However, this approach
requires a complete response variable for training the forest.

Instead, we directly predict the missing values using a random forest
trained on the observed parts of the data set. For an arbitrary variable
${\bf X}_s$ including missing values at entries ${\bf i}^{(s)}_{mis}
\subseteq \{1,\dots, n\}$ we can
separate the data set in four parts:
\begin{enumerate}
\item The observed values of variable ${\bf X}_s$, denoted by ${\bf
    y}^{(s)}_{obs}$;
\item the missing values of variable ${\bf X}_s$, denoted by ${\bf
    y}^{(s)}_{mis}$;
\item the variables other than ${\bf X}_s$ with observations
  ${\bf i}^{(s)}_{obs}=\{1,\dots,n\}\setminus{\bf i}^{(s)}_{mis}$ denoted by ${\bf x}^{(s)}_{obs}$;
\item the variables other than ${\bf X}_s$ with observations ${\bf
    i}^{(s)}_{mis}$ denoted by ${\bf x}^{(s)}_{mis}$.
\end{enumerate}
Note that ${\bf x}^{(s)}_{obs}$ is typically not completely observed since
the index ${\bf i}^{(s)}_{obs}$ corresponds to the observed values of the
variable ${\bf X}_s$. Likewise, ${\bf x}^{(s)}_{mis}$ is typically not
completely missing.

To begin, make an initial guess for the missing values in ${\bf X}$ using
mean imputation or another imputation method. Then, sort the variables
${\bf X}_s, s=1,\dots,p$ according to the amount of missing values starting
with the lowest amount. For each variable ${\bf X}_s$ the missing values
are imputed by first fitting a random forest with response ${\bf
  y}^{(s)}_{obs}$ and predictors ${\bf x}^{(s)}_{obs}$; then, predicting
the missing values ${\bf y}^{(s)}_{mis}$ by applying the trained random
forest to ${\bf x}^{(s)}_{mis}$. The imputation procedure is repeated until
a stopping criterion is met. The pseudo algorithm \ref{mf.alg} gives a
representation of the missForest method.

\begin{algorithm}[H]
  \caption{Impute missing values with random forest.}
  \label{mf.alg}
  \begin{algorithmic}[1]
    \REQUIRE ${\bf X}$ an $n \times p$ matrix, stopping criterion
    $\gamma$
    \STATE Make initial guess for missing values;
    \STATE ${\bf k} \leftarrow$ vector of sorted indices
    of columns in ${\bf X}$\\w.r.t. increasing amount of missing values;
    \WHILE{not $\gamma$}
    \STATE ${\bf X}^{imp}_{old} \leftarrow$ store previously imputed matrix;
    \FOR{$s$ in ${\bf k}$}
    \STATE Fit a random forest: ${\bf y}^{(s)}_{obs} \sim {\bf
      x}^{(s)}_{obs}$;
    \STATE Predict ${\bf y}^{(s)}_{mis}$ using ${\bf x}^{(s)}_{mis}$;
    \STATE ${\bf X}^{imp}_{new} \leftarrow$ update imputed matrix,
    using predicted ${\bf y}^{(s)}_{mis}$;
    \ENDFOR
    \STATE update $\gamma$.
    \ENDWHILE
    \RETURN the imputed matrix ${\bf X}^{imp}$
  \end{algorithmic}
\end{algorithm}

The stopping criterion $\gamma$ is met as soon as the difference between
the newly imputed data matrix and the previous one increases for the first
time with respect to both variable types, if present. Here, the difference
for the set of continuous variables ${\bf N}$ is defined as
\[
\Delta_{N}=\frac{\sum_{j\in {\bf N}}({\bf X}^{imp}_{new} - {\bf
    X}^{imp}_{old})^2}{\sum_{j\in {\bf N}}({\bf X}^{imp}_{new})^2},
\]
and for the set of categorical variables ${\bf F}$ as
\[
\Delta_{F}=\frac{\sum_{j\in{\bf F}}\sum_{i=1}^{n}{\bf I}_{{\bf X}^{imp}_{new} \neq {\bf
      X}^{imp}_{old}}}{\#\textrm{NA}},
\]
where $\#\textrm{NA}$ is the number of missing values in the categorical
variables.

After imputing the missing values the performance is
assessed using the normalised root mean squared error (NRMSE, \cite{oba03})
for the continuous variables which is defined by
\[
\textrm{NRMSE} = \sqrt{\frac{\textrm{mean}\left(({\bf X}^{true}-{\bf
        X}^{imp})^2\right)}{\textrm{var}\left({\bf X}^{true}\right)}},
\]
where ${\bf X}^{true}$ is the complete data matrix and ${\bf X}^{imp}$ the
imputed data matrix. We use mean and var as short notation for empirical
mean and variance computed over the continuous missing values only. For
categorical variables we use the proportion of falsely classified entries
(PFC) over the categorical missing values, $\Delta_{F}$. In both cases good
performance leads to a value close to 0 and bad performance to a value
around 1.

When a RF is fit to the observed part of a variable we also get an OOB
error estimate for that variable. After the stopping criterion $\gamma$ was
met we average over the set of variables of the same type to approximate
the true imputation errors. We assess the performance of this estimation by
comparing the absolute difference between true imputation error and OOB
imputation error estimate in all simulation runs.

  \section{Methods}\label{methods.sec} 
  We compare missForest with four methods on ten different data sets where
  we distinguish between situations with continuous variables only,
  categorical variables only and mixed variable types.
  
  The most well-known method for imputation of continuous data sets
  especially in the field of gene expression analysis is the KNNimpute
  algorithm by \cite{troyanskaya01}. A missing value variable ${\bf
    X}_j$ is imputed by finding its $k$ nearest observed variables and
  taking a weighted mean of these $k$ variables for imputation. Thereby,
  the weights depend on the distance of the variable ${\bf X}_{j}$. The
  distance itself is usually chosen to be the Euclidean distance.

  When using KNNimpute the choice of the tuning parameter
  $k$ can have a large effect on the performance of the imputation. However,
  this parameter is not known beforehand. Since our method includes no such
  parameter we implement a cross-validation (see Algorithm \ref{adaptKNN})
  to obtain a suitable $k$.

  \begin{algorithm}[H]
    \caption{Cross-validation KNN imputation.}
    \label{adaptKNN}
    \begin{algorithmic}[1]
      \REQUIRE ${\bf X}$ an $n \times p$ matrix, number of validation sets
      $l$, range of suitable number of nearest neighbours ${\bf K}$
      \STATE ${\bf X}^{CV} \leftarrow$ initial imputation using mean imputation;
      \FOR{$t$ in $1, \dots, l$}
      \STATE ${\bf X}^{CV}_{mis,t} \leftarrow$ artificially introduce
      missing values to ${\bf X}^{CV}$;
      \FOR{$k$ in ${\bf K}$}
      \STATE ${\bf X}^{CV}_{KNN,t} \leftarrow$ KNN imputation of ${\bf
        X}^{CV}_{mis,t}$ using $k$ nearest neighbours;
      \STATE $\varepsilon_{k, t} \leftarrow$ error of KNN imputation for
      $k$ and $t$;
      \ENDFOR
      \ENDFOR
      \STATE $k_{best} \leftarrow
      \underset{k}{\operatorname{argmin}}\frac{1}{l}\sum_{t=1}^{l}\varepsilon_{k,t}$;
      \STATE ${\bf X}^{imp} \leftarrow$ KNN imputation of ${\bf X}$ using
      $k_{best}$ nearest neighbours.
    \end{algorithmic}
  \end{algorithm}

  In the original paper of \cite{troyanskaya01} the data was not
  standardized before applying the KNNimpute algorithm. This constitutes no
  issue in the case of gene expression data because such data generally
  consists of variables on similar scales. However, we are applying the
  KNNimpute algorithm to data sets with varying scales in the variables. To
  avoid variance based weighting of the variables we scale them to a unit
  standard deviation. We also center the variables at zero. After
  imputation the data is retransformed such that the error is computed on
  the original scales. This last step is performed because missForest does
  not need any transformation of the data and we want to compare the
  performance of the methods on the original scales of the data.

  Another approach for continuous data, especially in the case of
  high-dimensional normal data matrices, is presented by \cite{staedler10b}
  using an EM-type algorithm. In their Missingness Pattern Alternating
  Imputation and $l_1$-penalty (MissPALasso) algorithm the missing
  variables are regressed on the observed ones using the lasso penalty by
  \cite{tibshirani96}. In the following E step the obtained regression
  coefficients are used to partially update the latent distribution. The
  MissPALasso has also a tuning parameter $\lambda$ for the penalty. As
  with KNNimpute we use cross-validation to tune $\lambda$ (cf. Algorithm
  \ref{adaptKNN}). When applying MissPALasso the data is standardized as
  regularization with a single $\lambda$ requires the different regressions
  to be on the same scale.\\

  In the comparative experiments with categorical or mixed-type variables
  we use the MICE algorithm by \cite{vanbuuren99} based on the multivariate
  multiple imputation scheme of \cite{schafer97}. In contrast to
    the latter the conditional distribution for the missing data in each
    incomplete variable is specified in MICE, a feature called fully
    conditional specification by \cite{vanbuuren07}. However, the existence
    of a multivariate distribution from which the conditional distribution
    can be easily derived is assumed. Furthermore, iterative Gibbs sampling
    from the conditional distributions can generate draws from the
    multivariate distribution. We want to point out that MICE in its
  default setup is not mainly intended for simple missing value
  imputation. Using the multiple imputation scheme MICE allows for
  assessing the uncertainty of the imputed values. It includes
    features to pool multiple imputations, choose individual sampling
    procedures and allows for passive imputation controlling the sync of
    transformed variables. In our experiments we used MICE with either
  linear regression with normal errors or mean imputation for continuous
  variables, logistic regression for binary variables and polytomous
  logistic regression for categorical variables with more than two
  categories.

  For comparison across different types of variables we apply the KNNimpute
  algorithm with dummy coding for the categorical variables. This is done
  by coding a categorical variable ${\bf X}_{j}$ into $m$ dichotomous
  variables $\tilde{{\bf X}}_{j,m} \in \{-1,1\}$. Application of the
  KNNimpute algorithm for categorical data can be summarized as:
  \begin{enumerate}
    \item Code all categorical variables into $\{-1,1\}$-dummy variables;
    \item standardize all variables to mean 0 and standard deviation 1;
    \item apply the cross-validated KNNimpute method from Algorithm
      \ref{adaptKNN};
    \item retransform the imputed data matrix to the original scales;
    \item code the dummy variables back to categorical variables;
    \item computed the imputation error.
  \end{enumerate}

  For each experiment we perform 50 independent simulations where 10\%, 20\% or
  30\% of the values are removed completely at random. Each method is then
  applied and the NRMSE, the PFC or both are computed (see Section
  \ref{approach.sec}). We perform a paired Wilcoxon test of the
    error rates of the compared methods versus the error rates of
    missForest. In addition, the OOB error estimates of missForest is
  recorded in each simulation.

\section{Results}\label{result.sec}
\subsection{Continuous variables only}
First, we focus on continuous data. We investigate the following four
publicly available data sets:
\begin{itemize}
\item {\bf Isoprenoid gene network in {\it A. thaliana}}: This gene network
  includes $p=39$ genes each with $n=118$ gene expression profiles
  corresponding to different experimental conditions. For more details on
  this data set see \cite{wille04}.
\item {\bf Voice measures in Parkinson's patients}: The data described by
  \cite{little08} contains a range of biomedical voice measurements from
  31 individuals, 23 with Parkinson's disease (PD). There are $p=22$
  particular voice measurements and $n=195$ voice recordings from these
  individuals. The data set also contains a response variable giving the
  health status. Dealing only with continuous variables the response was
  removed from the data. We will return to this later on.
\item {\bf Shapes of musk molecules}: This data set describes 92 molecules of
  which 47 are musks and 45 are non-musks. For each molecule $p=166$ features
  describe its conformation, but since a molecule can have many
  conformations due to rotating bonds, there are $n=476$ different
  low-energy conformations in the set. The classification into musk and
  non-musk molecules is removed.
\item {\bf Insulin gene expression}: This high-dimensional data set originates
  from an analysis by \cite{wu07} of {\it vastus lateralis} muscle biopsies
  from three different types of patients following insulin treatment. The
  three types are insulin-sensitive, insulin-resistant and diabetic
  patients. The analysis involves $p=12'626$ genes whose expression levels
  were measured from $n=110$ muscle biopsies. Due to computation time we
  only perform 10 simulations instead of 50.
\end{itemize}

\begin{figure}
  \centering
  \includegraphics[width=70mm]{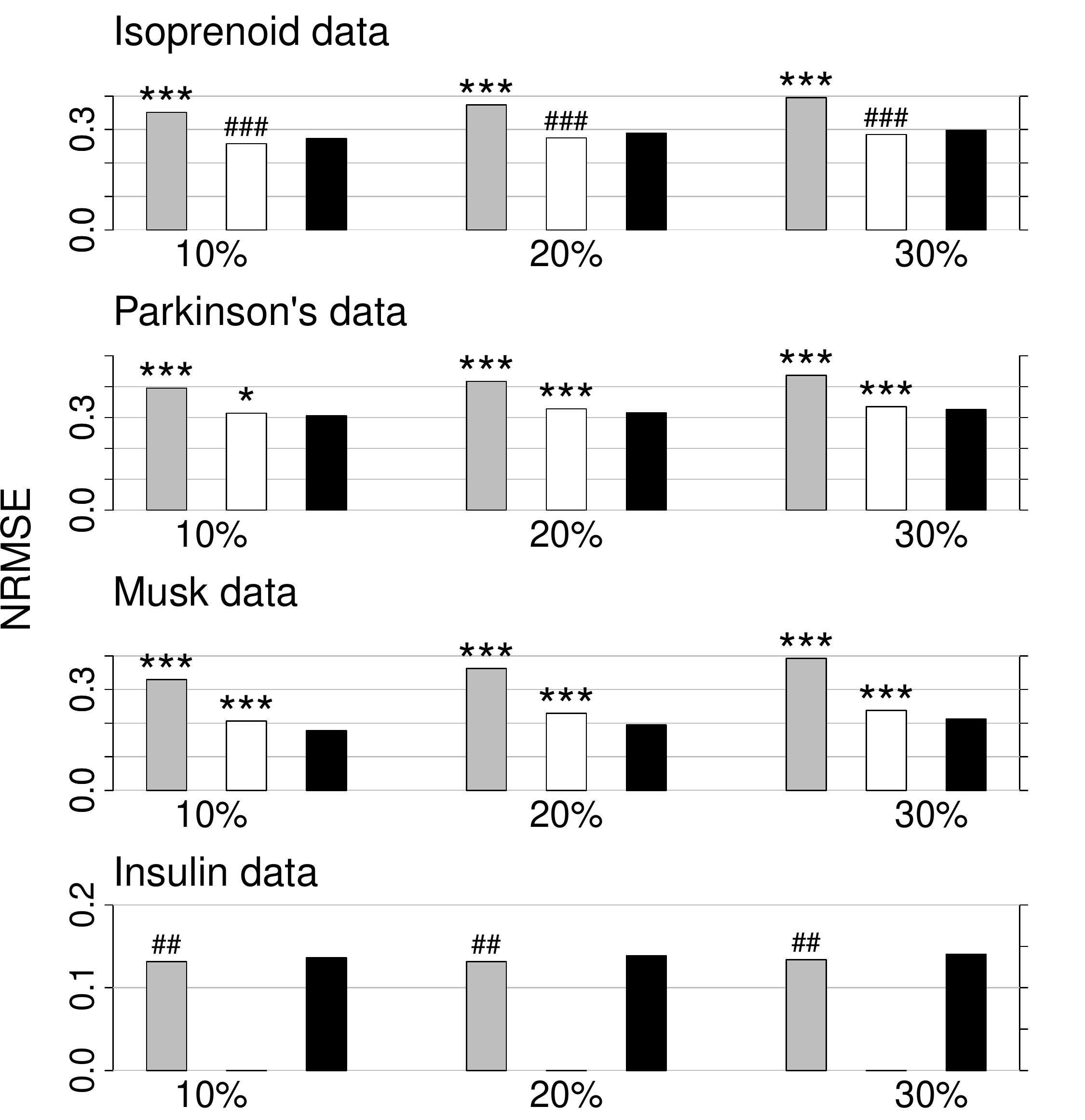}
  \caption{{\bf Continuous data.} Average NRMSE for KNNimpute (grey),
    MissPALasso (white) and missForest (black) on four different data sets
    and three different amounts of missing values, i.e., 10\%, 20\% and
    30\%. Standard errors are in the order of magnitude of
    $10^{-4}$. Significance levels for the paired Wilcoxon tests in favour
    of missForest are encoded as ``*'' $<$0.05, ``**'' $<$0.01 and ``***''
    $<$0.001. If the average error of the compared method is smaller than
    that of missForest the significance level is encoded by a hash (\#)
    instead of an asterisk. In the lowermost data set results for
    MissPALasso are missing due to the methods limited capability with
    regard to high dimensions.}\label{cont.fig}
\end{figure}

Results are given in Figure \ref{cont.fig}. We can see that missForest
performs well, sometimes reducing the average NRMSE by up to 25\% with
respect to KNNimpute. In case of the musk molecules data the reduction is
even above 50\%. The MissPALasso performs slightly better than missForest
on the gene expression data. However, there are no results for the
MissPALasso in case of the Insulin data set because the high dimension makes computation not feasible.

For continuous data the missForest algorithm typically reaches the stopping
criterion quite fast needing about 5 iterations. The imputation takes about
10 times as long as performing the cross-validated KNNimpute where
$\{1,\dots,15\}$ is the set of possible numbers of neighbours. For the
Insulin data set an imputation takes on average 2 hours on a customary
available desktop computer.

\subsection{Categorical variables only}
We also consider data sets with only categorical variables. Here, we use
the MICE algorithm described in Section \ref{methods.sec} instead of the
MissPALasso. We use a dummy implementation of the KNNimpute algorithm to
deal with categorical variables (see Section \ref{methods.sec}). We apply
the methods to the following data sets:

\begin{itemize}
\item {\bf Cardiac single photon emission computed tomography (SPECT) images}:
  \cite{kurgan01} discuss this processed data set summarizing over 3000 2D
  SPECT images from $n=267$ patients in $p=22$ binary feature patterns.
\item {\bf Promoter gene sequences in {\it E. coli}}: The data set contains
  sequences found by \cite{harley87} for promoters and sequences found by
  \cite{towell90} for non-promoters totalling $n=106$. For each candidate a
  sequence of 57 base pairs was recorded. Each variable can take one of
  four DNA nucleotides, i.e., adenine, thymine, guanine or cytosine. Another
  variable distinguishes between promoter and non-promoter instances.
\item {\bf Lymphography domain data}: The observations were obtained from
  patients suffering from cancer in the lymphatic of the immune system. For
  each of the $n=148$ lymphoma $p=19$ different properties were recorded
  mainly in a nominal fashion. There are nine binary variables. The rest
  of the variables have three or more levels.
\end{itemize}

\begin{figure}
  \centering
  \includegraphics[width=70mm]{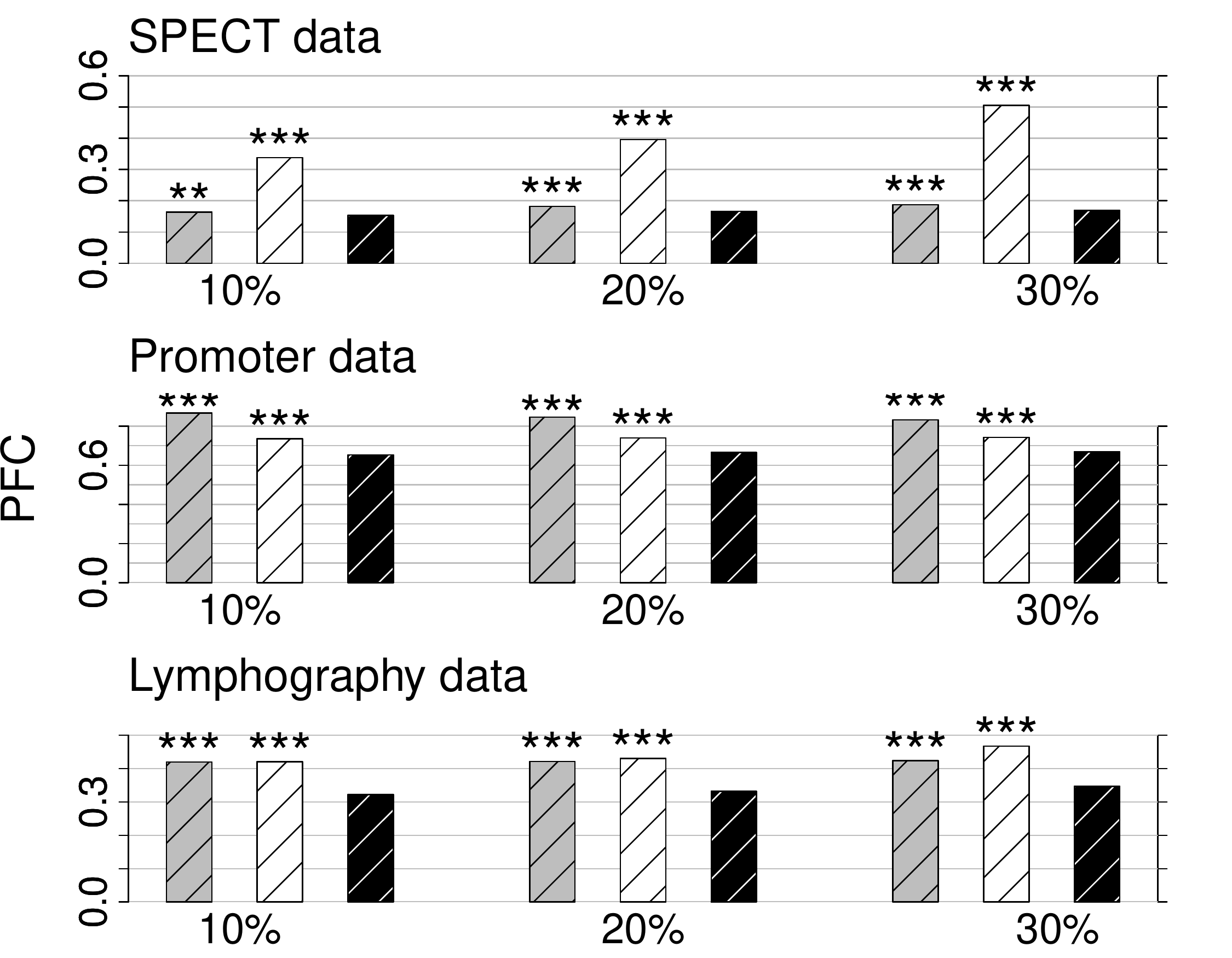}
  \caption{{\bf Categorical data.} Average PFC for cross-validated
    KNNimpute (grey), MICE (white) and missForest (black) on three
    different data sets and three different amounts of missing values,
    i.e., 10\%, 20\% and 30\%. Standard errors are in the order of
    magnitude of $10^{-4}$. Significance levels for the paired Wilcoxon
    tests in favour of missForest are encoded as ``*'' $<$0.05, ``**''
    $<$0.01 and ``***'' $<$0.001.}\label{cat.fig}
\end{figure}

In Figure \ref{cat.fig} we can see that missForest is always imputing the
missing values better than the compared methods. In some cases -- namely
for the SPECT data -- the decrease of PFC compared to MICE is up to
60\%. However, for the other data sets the decrease is less pronounced
ranging around 10 - 20\% -- but there still is a decrease. The amount of
missing values on the other hand seems to have only a minor influence on
the performance of all methods. Except for MICE on the SPECT data, error
rates remain almost constant increasing only by 1 - 2\%. We pointed out
earlier that MICE is not primarily tailored for imputation performance but
offers additional possibilities of assessing uncertainty of the imputed
values due to the multiple imputation scheme. Anyhow, the results using the
cross-validated KNNimpute (see Algorithm \ref{adaptKNN}) on the dummy-coded
categorical variables is surprising. The imputation for missForest needs on
average 5 times as long as a cross-validated imputation using KNNimpute.

\subsection{Mixed-type variables}\label{mixed:ssec}
In the following we investigate four data sets where the first one has
already been introduced, i.e.\ {\it musk molecules} data including the
categorical response yielding the classification. The other data sets are:
\begin{itemize}
\item {\bf Proteomics biomarkers for Gaucher's disease}: Gaucher's
  disease is a rare inherited enzyme deficiency. In this data set
  \cite{smit07} present protein arrays for biomarkers ($p=590$) from
  blood serum samples ($n=40$). The binary response distinguishes between
  disease status.
\item {\bf Gene finding over prediction (GFOP) peptide search}: This data set
  comprises mass-spectrometric measurements of $n=595$ peptides from two
  shotgun proteomics experiments on the nematode {\it Caenorhabditis
    elegans}. The collection of $p=18$ biological, technical and analytical
  variables had the aim of novel peptide detection in a search on an extended
  database using established gene prediction methods.
\item {\bf Children's Hospital data}: This data set is the product of a
  systematic long-term review of children with congenital heart defects
  after open-heart surgery. Next to defect and surgery related variables
  also long-term psychological adjustment and health-related quality of
  life was assessed. After removing observations with missing values the
  data set consists of $n=55$ patients and $p=124$ variables of which 48
  are continuous and 76 are categorical. For further details see
  \cite{latal09}.
\end{itemize}

\begin{figure}
  \centering
  \includegraphics[width=70mm]{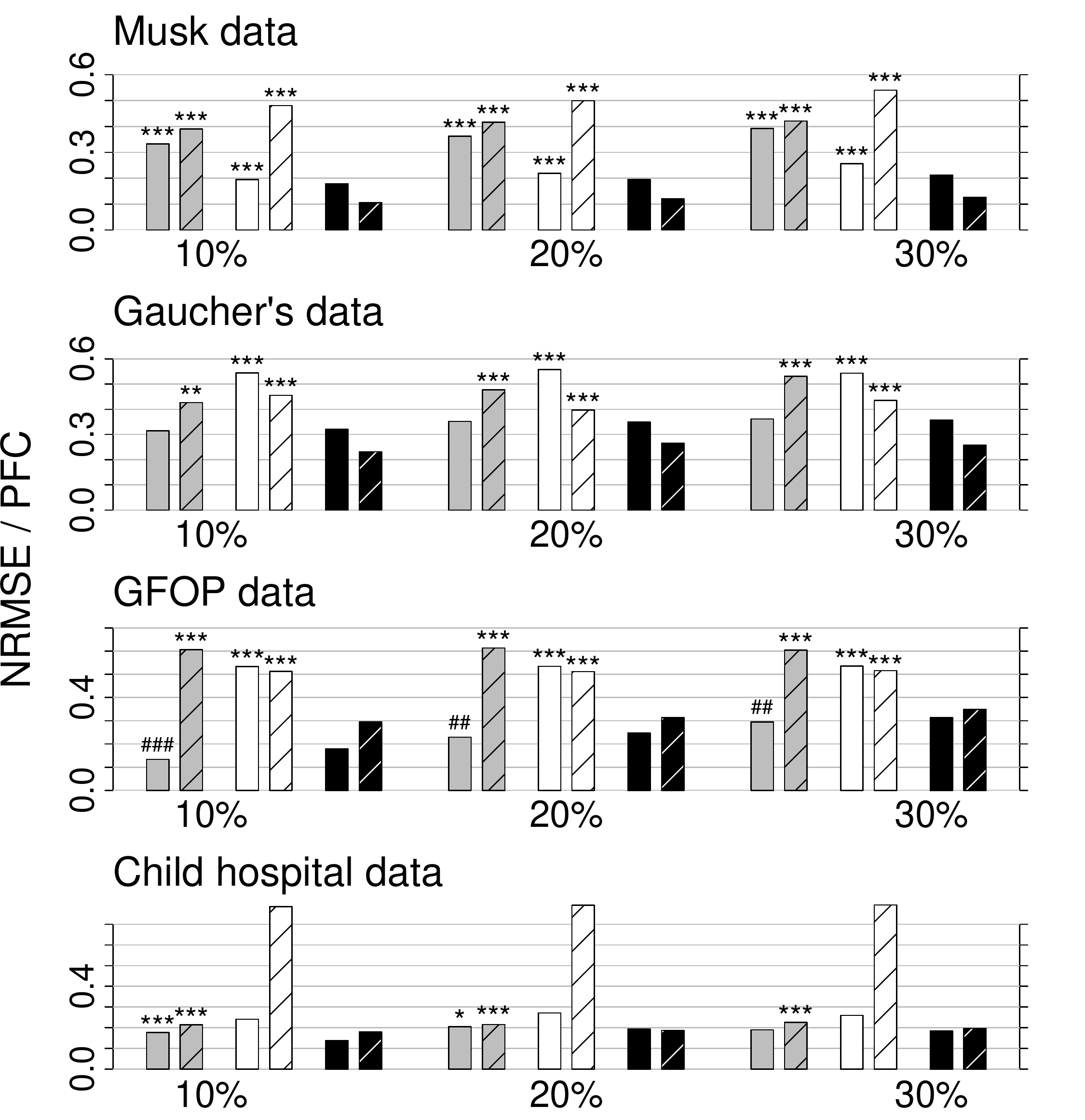}
  \caption{{\bf Mixed-type data.} Average NRMSE (left bar) and PFC (right
    bar, shaded) for KNNimpute (grey), MICE (white) and missForest (black)
    on four different data sets and three different amounts of missing
    values, i.e., 10\%, 20\% and 30\%. Standard errors are in the order of
    magnitude of $10^{-3}$. Significance levels for the paired Wilcoxon
    tests in favour of missForest are encoded as ``*'' $<$0.05, ``**''
    $<$0.01 and ``***'' $<$0.001. If the average error of the compared
    method is smaller than that of missForest the significance level is
    encoded by a hash (\#) instead of an asterisk. Note that, due to
    ill-distribution and near dependence in the Child hospital data, the
    results for MICE have to be treated with caution (see Section
    \ref{mixed:ssec}).}\label{mixed.fig}
\end{figure}

The results of this comparison are given in Figure \ref{mixed.fig}. We can
see that missForest performs better than the other two methods, again
reducing imputation error in many cases by more than 50\%. For the GFOP
data, KNNimpute has a slightly smaller NRMSE than missForest but makes
twice as much error on the categorical variables. Generally, with respect
to the amount of missing values the NRMSE tends to have a greater
variability than the PFC which remains largely the same.  

The imputation results for MICE on the Children's Hospital data
  have to be treated cautiously. Since this data set contains
  ill-distributed and nearly dependent variables, e.g., binary variables
  with very few observations in one category, the missingness pattern has a
  direct influence on the operability of the MICE implementation in the
  statistical software R. The imputation error illustrated in Figure
  \ref{mixed.fig} was computed from 50 successful simulations by randomly
  generating missingness patterns, which did not include only complete
  cases or no complete cases at all within the categories of the
  variables. Therefore, the actual numbers of simulations were larger than
  50 for all three missing value amounts. Furthermore, nearly dependent
  variables were removed after each introduction of missing values. This
  leads to an average of 7 removed variables in each simulation. Due to
  this ad-hoc manipulation for making the MICE implementation work, we do
  not report significance statements for the imputation error.

\subsection{Estimating imputation error}
\begin{figure}[tb]
  \centerline{\includegraphics[width=45mm]{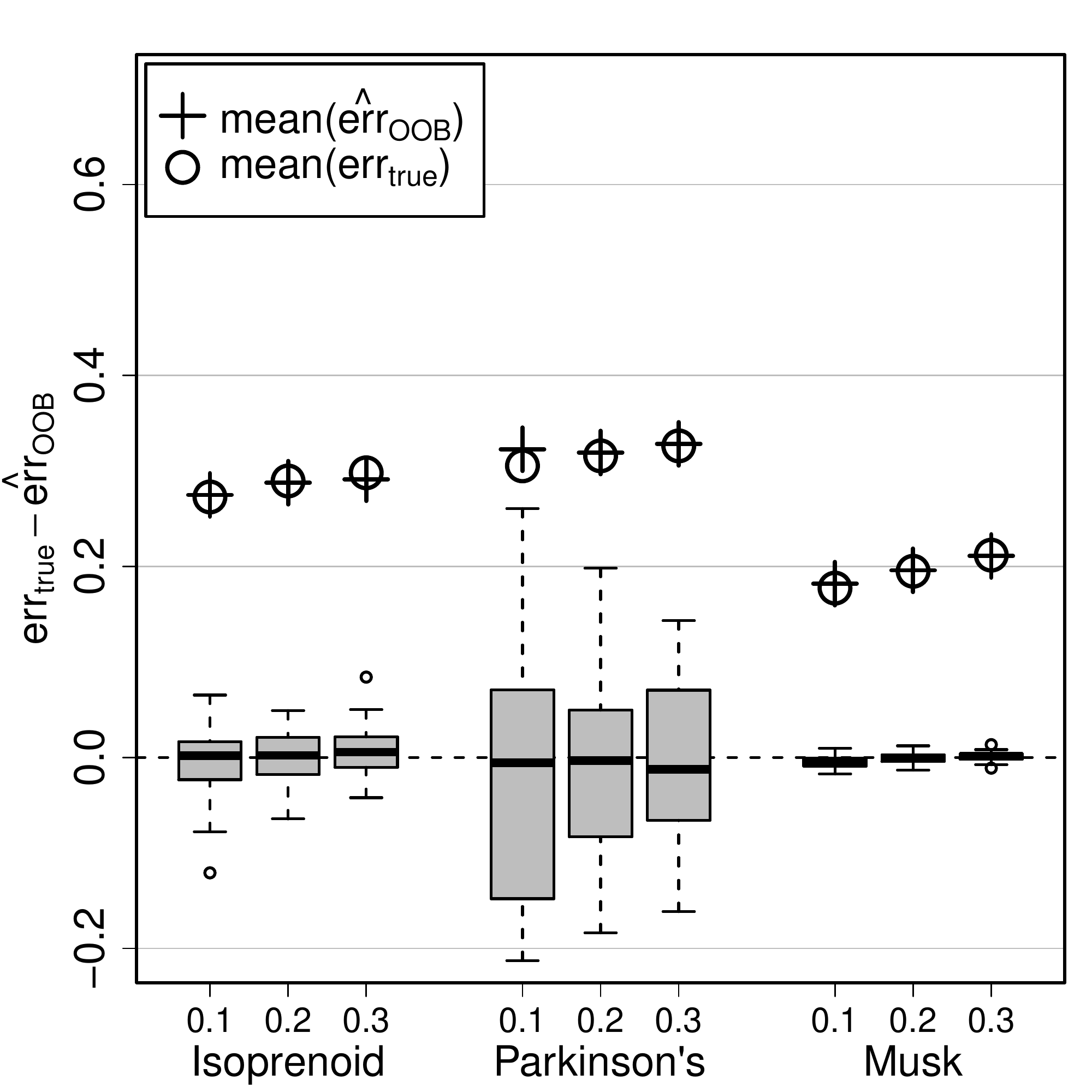}
              \includegraphics[width=45mm]{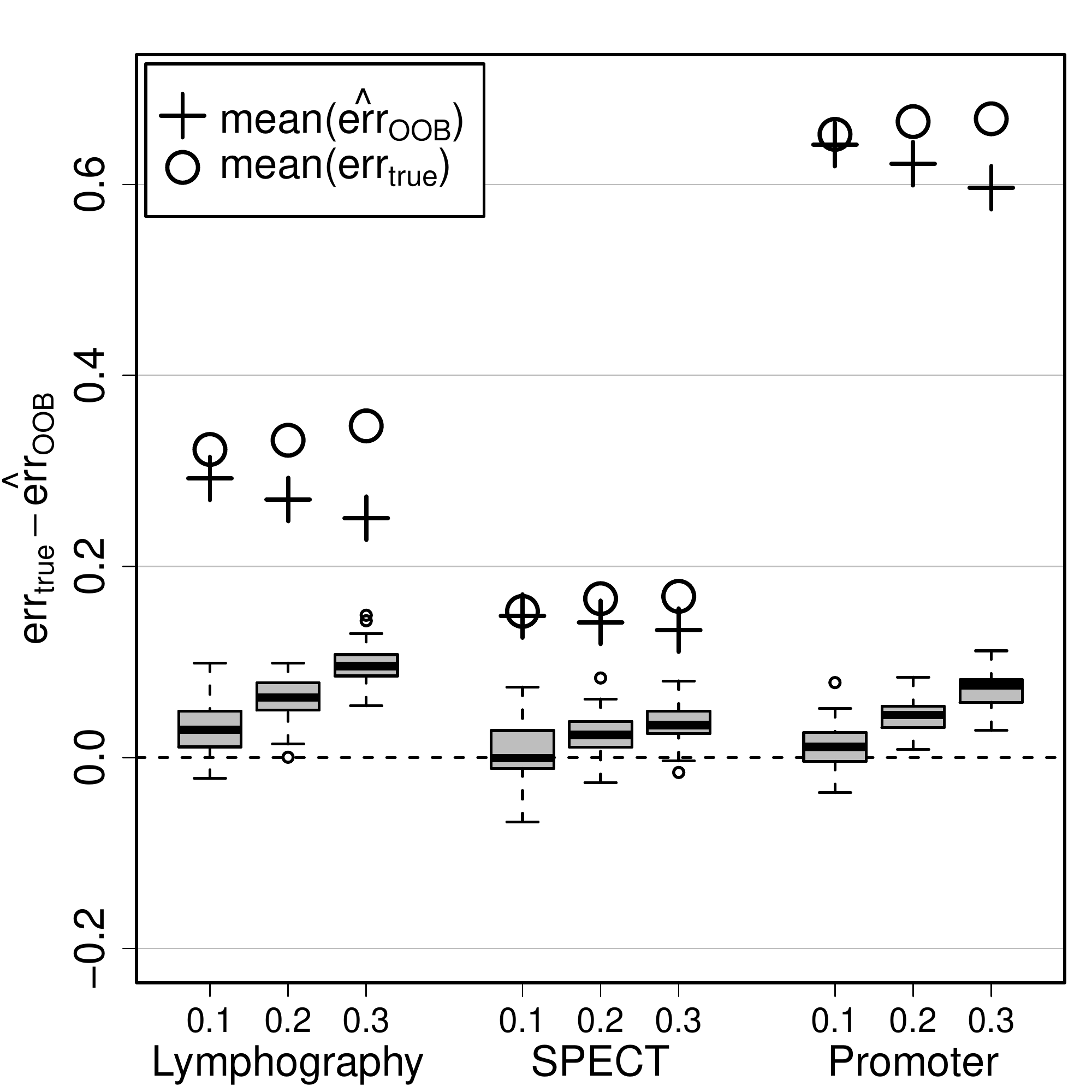}}
            \caption{Difference of true imputation error
              $\textrm{err}_{\textrm{true}}$ and OOB imputation error
              estimate $\widehat{\textrm{err}}_{\textrm{OOB}}$ for the
              continuous data sets (left) and the categorical data sets
              (right) and three different amounts of missing values,
              i.e., 0.1, 0.2 and 0.3. In each case the average
              $\textrm{err}_{\textrm{true}}$ (circle) and the average
              $\widehat{\textrm{err}}_{\textrm{OOB}}$ (plus) over all simulations
              is given.}\label{OOB.fig}
\end{figure}

In each experiment we get for each simulation run an OOB estimate for the
imputation error. In Figure \ref{OOB.fig} the differences of true
imputation error, $\textrm{err}_{\textrm{true}}$, and OOB error estimates,
$\widehat{\textrm{err}}_{\textrm{OOB}}$, are illustrated for the
continuous and the categorical data sets. Also, the mean of the true
imputation error and the OOB error estimate over all simulations is
depicted.

We can see that for the Isoprenoid and Musk data sets the OOB estimates are
very accurate only differing from the true imputation error by a few
percents. In case of the Parkinson's data set the OOB estimates exhibit a
lot more variability than in all other data sets. However, on average the
estimation is comparably good.  For the categorical data sets the
estimation accuracy behaves similarly over all scenarios. The OOB estimates
tend to underestimate the imputation error with increasing amount of
missing values.  Apparently, the absolute size of the imputation error
seems to play a minor role in the accuracy of the OOB estimates which can
be seen nicely when comparing the SPECT and the Promoter data.

\subsection{Computational efficiency}
We assess the computational cost of missForest by comparing the runtimes of
imputation on the previous data sets. Table \ref{runtime.tab} shows the
runtimes in seconds of all methods on the analyzed data sets. We can see
that KNNimpute is by far the fastest method. However, missForest runs
considerably faster than MICE and the MissPALasso. In addition, applying
missForest did not require antecedent standardization of the data,
laborious dummy coding of categorical variables nor implementation of CV
choices for tuning parameters.
\begin{table}[h]
  \centering
  {\begin{tabular}{lcccccc}\hline
      Data set     &  n  &  p  & KNN   & MissPALasso & MICE & missForest \\
      \hline
      Isoprenoid   & 118 &  39 & 0.8 & 170  & -    & 5.8  \\
      Parkinson's  & 195 &  22 & 0.7 & 120  & -    & 6.1  \\
      Musk (cont.) & 476 & 166 & 13  & 1400 & -    & 250  \\
      Insulin      & 110 &12626& 1800& n/a  & -    & 6200 \\
      \hline
      SPECT        & 267 &  22 & 1.3 &   -  & 37   & 5.5  \\
      Promoter     & 106 &  57 & 14  &   -  & 4400 & 38   \\
      Lymphography & 148 &  19 & 1.1 &   -  & 93   & 7.0  \\
      \hline
      Musk (mixed) & 476 & 167 & 27  & -    & 2800 & 500  \\
      Gaucher's    &  40 & 590 & 1.3 & -    & 130  & 29   \\
      GFOP         & 595 &  18 & 2.7 & -    & 1400 & 40   \\
      Children     &  55 & 124 & 2.7 & -    & 4000 & 110  \\
      \hline
    \end{tabular}}
  \caption{Average runtimes [s] for imputing the analyzed data
    sets. Runtimes are averaged over the amount of missing 
    values since this has a negligible effect on computing
    time.}\label{runtime.tab}
\end{table}

There are two possible ways to speed up computation. The first one is to
reduce the number of trees grown in each forest. In all comparative studies
the number of trees was set to 100 which offers high precision but
increased runtime. In Table \ref{tree.tab} we can see that changing the
number of trees in the forest has a stagnating influence on imputation
error, but a strong influence on computation time which is approximately
linear in the number of trees.

The second one is to reduce the number of variables randomly selected at
each node ($m_{\textrm{try}}$) to set up the split. Table \ref{tree.tab}
shows that increasing $m_{\textrm{try}}$ has limited effect on imputation
error, but computation time is strongly increased. Note that for
$m_{\textrm{try}}=1$ we do not longer have a random forest, since there is
no more choice between variables to split on. This leads to a much higher
imputation error, especially for the cases with low numbers of bootstrapped
trees. We use for all experiments $\lfloor\sqrt{p}\rfloor$ as default
value, e.g., in the GFOP data this equals 4.

\begin{table}[tb]
  \centering
  {\begin{tabular}{cccccccc}\hline
      $m_{\textrm{try}}$&&\multicolumn{5}{c}{$n_{\textrm{tree}}$}\\\cline{3-7}
      && 10        & 50       & 100       & 250       & 500      \\
      \hline
      \multirow{2}{*}{1} && 36.8/35.5 & 27.4/32.3 & 20.4/31.3 & 17.2/30.0 &
      16.0/30.8\\
      && 2.5s      & 3.2s      & 3.9s      & 5.8s      & 9.2s\\
      && & & & & \\
      \multirow{2}{*}{2} && 34.9/31.8 & 24.8/29.2 & 18.3/28.8 & 16.0/28.6 & 15.5/29.1\\
      && 6.9s      & 11.8s     & 15.0s     & 25.2s     & 39.3s\\
      && & & & & \\
      \multirow{2}{*}{4} && 34.9/31.3 & 24.4/28.9 & 17.9/28.2 & 15.4/28.2 & 15.8/28.7\\
      && 16.5s     & 25.1s     & 35.0s     & 49.0s     & 83.3s\\
      && & & & & \\
      \multirow{2}{*}{8} && 34.7/31.4 & 24.3/28.9 & 18.1/27.8 & 15.2/27.8 & 15.7/28.6\\
      && 39.2s     & 57.4s     & 84.4s     & 130.2s    & 190.8s\\
      && & & & & \\
      \multirow{2}{*}{16}&& 34.6/30.9 & 24.3/28.7 & 18.1/28.0 & 15.4/27.8 & 15.6/28.5\\
      && 68.7s     & 99.7s     & 172.2s    & 237.6s    & 400.7s\\
      \hline
    \end{tabular}}
  \caption{Average imputation error (NRMSE/PFC in percent)
    and runtime (in seconds) with different numbers of trees
    ($n_{\textrm{tree}}$) grown in each forest and variables tried
    ($m_{\textrm{try}}$) at each node of the trees. Here, we consider the
    GFOP data set with artificially introduced 10\% of missing
    values. For each comparison 50 simulation runs were performed using
    always the same missing value matrix for all numbers of
    trees/randomly selected variables for a
    single simulation.}\label{tree.tab}
\end{table}

\section{Conclusion}
Our new algorithm, missForest, allows for missing value imputation on
basically any kind of data. In particular, it can handle multivariate data
consisting of continuous and categorical variables
simultaneously. MissForest has no need for tuning parameters nor does it
require assumptions about distributional aspects of the data. We show on
several real data sets coming from different biological and medical fields
that missForest outperforms established imputation methods like $k$-nearest
neighbours imputation or multivariate imputation using chained
equations. Using our OOB imputation error estimates missForest offers a way
to assess the quality of an imputation without the need of setting aside
test data nor performing laborious cross-validations. For subsequent
analysis these error estimates represent a mean of informal reliability
check for each variable. The full potential of missForest is deployed when
the data includes complex interactions or nonlinear relations between
variables of unequal scales and different type. Furthermore, missForest can
be applied to high-dimensional data sets where the number of variables may
greatly exceed the number of observations to a large extent and still
provides excellent imputation results.

\section*{Acknowledgement}
Except for the Isoprenoid, the Lymphography, the Children's Hospital and
the GFOP data all other data sets were obtained from the UCI machine
learning repository (\cite{UCI10}). The GFOP data set was obtained from the
Institute of Molecular Systems Biology, Zurich, Switzerland. Thanks go to
L.~Reiter for providing the data. The Lymphography data set was obtained
from the University Medical Centre, Institute of Oncology, Ljubljana,
Slovenia. Thanks go to M.~Zwitter and M.~Soklic for providing the data. The
Children's Hospital data set was obtained from the Child Development Center
at the University Children's Hospital Zürich, Switzerland. Thanks go to
B.~Latal and I.~Beck for providing the data. Finally, we thank two
anonymous referees for their constructive comments.

\bibliographystyle{natbib}

\bibliography{myBib}

\end{document}